\documentclass[aps,pre,twocolumn,superscriptaddress,showpacs,showkeys ]{revtex4-1}

\usepackage{graphicx}                   
\usepackage{hyperref}                   
\usepackage{amsmath,amssymb}            
\usepackage{epstopdf}
\usepackage{subcaption}					

\usepackage{color}
\definecolor{orange}{rgb}{1,0.5,0}
\definecolor{brown}{rgb}{0.65, 0.16, 0.16}
\definecolor{phlox}{rgb}{0.87, 0.0, 1.0}

\graphicspath{{figs/}}					

\begin{document}

\title{Avalanches on the Complex Network of Rigan Earthquake, Virtual Seismometer Technique,\\
	 Criticality and Seismic Cycle}

\author{M. N. Najafi}
\affiliation{Department of Physics, University of Mohaghegh Ardabili, P.O. Box 179, Ardabil, Iran}
\email{morteza.nattagh@gmail.com}

\author{M. Rahimi-Majd}
\affiliation{Department of Physics, Shahid Beheshti University, Velenjak, Tehran 19839, Iran}
\affiliation{ Ibn-Sina Multidisciplinary laboratory, Department of Physics, Shahid Beheshti University, Velenjak, Tehran 19839, Iran}
\email{rahimimajd.milad@gmail.com}

\author{T. Shirzad}
\affiliation{Institute of Astronomy, Geophysics and Atmospheric Sciences, University of Sao Paulo, 05508-090, Sao Paulo, Brazil. t.shirzad@iag.usp.br ; ORCID: 0000-0002-8382-4990}
\email{taghishirzadiraj@gmail.com}

\begin{abstract}
We base our study on the statistical analysis of the Rigan earthquake 2010 December 20, which consists of estimating the earthquake network by means of virtual seismometer technique, and also considering the avalanche-type dynamics on top of this complex network.The virtual seismometer complex network shows power-law degree distribution with the exponent $\gamma=2.3\pm 0.2$. Our findings show that the seismic activity is strongly intermittent, and have a \textit{cyclic shape} as is seen in the natural situations, which is main finding of this study. The branching ratio inside and between avalanches reveal that the system is at (or more precisely close to) the critical point with power-law behavior for the distribution function of the size and the mass and the duration of the avalanches, and with some scaling relations between these quantities. The critical exponent of the size of avalanches is $\tau_S=1.45\pm 0.02$. We find a considerable correlation between the dynamical Green function and the nodes centralities. 
\end{abstract}


\maketitle

\section{Introduction}

Two strategies are often taken for explaining the observations of earthquakes: the quenched-disorder based models ascribing the observations to the geometric and material irregularities in the earth, and the dynamical-instability models attributing the complexities to the stochastic forcing arising from the dynamic nonuniformities~\cite{carlson1994dynamics}. In the former, the power-laws observed in an earthquake is related to geometric features of the fault structure~\cite{kagan1987statistical}. Whether the earth is operating according to one of these schemes or in a hybrid one remains an open and fundamental problem. The studies on seismic activities belong to one of the following three main categories: the phenomenological (if one insists not to use full-empirical) models that are based purely on the natural observations~\cite{telesca2001intermittent,telesca2012analysis,telesca2016visibility,pasten2017time}, the dynamical models on pre-existing fault networks~\cite{huang1998precursors}, and the dynamical (on- or off-lattice self-organized) models with random dynamic forces~\cite{bak1989earthquakes,sornette1989self,olami1992self}. The basic assumption for the latter is the common belief that the earthquakes (the ones which occur in the upper $\sim$ten kilometers of the earth's crust) arise as a consequence of frictional instabilities that cause stress, accumulated by large-scale plate motions over periods of hundreds of years, to be relieved in sudden stick-slip events~\cite{huang1998precursors}. The spring-block model is a clear example of such models in which the spring strain spreads throughout the system by means of an avalanche-like dynamics, and in each local activity the stress is distributed isotropically between the closest neighbors~\cite{burridge1967model,brown1991simplified}. The present models, whatever they try to take the details of the dynamics of activities into account, suffer the lack of a lot of details arising from the complex nature of earth's crust. In this regard, the first and second categories which bring these details into the calculations as the background of their dynamical model~\cite{huang1998precursors} or so, work better.\\

The representation of seismic sequences as time series has been highly regarded in the literature as an efficient method to apply techniques derived from the nonlinear analysis. Using this, the basic properties of the system can be quantified in terms of e.g. scale-invariant correlations~\cite{descherevsky2003flicker}, $1/f$ noise~\cite{milotti20021}, or power-law decays~\cite{gutenberg1942earthquake}. These scale-invariant analysis and stochastic techniques are always preferable to ad hoc mathematical approaches since they deal with the real data as outlined above. These time series are composed of geophysical signals which are characterized by a spiky dynamic, with sudden and intense bursts of high frequency activity. The dynamics is an outcome of the rupture propagation with complex friction laws and barriers~\cite{carlson1994dynamics,ben1995slip,cochard1994dynamic}. From this point of view, visibility graph method~\cite{telesca2012analysis,telesca2016visibility} helps a lot to recognize the statistical features of the system. This method cannot tell us much about the dynamical features of the system. One may combine such complex-network supports (arising from the earth's crust activities) with avalanche-based (Self-organized critical) models to get closer to the real situations~\cite{bak1989earthquakes,sornette1989self,olami1992self}. One of the most popular methods for constructing these complex networks is by connecting two nodes (which are two main shocks) if they occur sequentially in the time sequence of the earthquakes. Apparently such methods suffer a crucial issue: two successive events are not necessarily causally related.\\

A feature of the time series of seismic process is the existence of sparse (low activity) temporal phases that are interspersed between those with relatively large density of the events, which can be viewed as the geometrical manifestation of intermittence~\cite{telesca2001intermittent,davis1994multifractal,telesca2009non}. The presence of intermittency in the spiky temporal dynamics in seismicity data reveals the effect of an heterogeneous lithosphere, taking place at many time scales~\cite{jaume1999evolving,huang1998precursors}. In Ref.~\cite{sammis1999seismic} the \textit{seismic cycle} during which a large event is followed by a shadow period of quiescence and then a new approach back toward the critical state, in which the events become larger is attributed to the large-scale heterogeneity. Here another possibility is presented that we which is much more like the second category, i.e. a dynamical avalanche-based model was defined on top of the complex network arising from the virtual seismometer technique for the Rigan earthquake on 2010 December 20. In our complex network, two nodes are related if their inter-event empirical Green's function (hereafter EGF) was satisfied the interferometric criteria. This technique was formulated by Curtis et al.,~\cite{curtis2009,galetti2012generalised} and suggested an alternative method to extract the inter-event EGF using the cross-correlating of event-pair synthetic waveforms recorded on a station~\cite{galetti2012generalised}. This method has further been developed by Shirzad~\cite{shirzad2019} on real data to study of hidden part of Kahurak fault plane. The dynamical model that we use is much like the bak-Tang-Weisendeld (BTW) sandpile model, which have already proved to be an acceptable choice for earthquakes~\cite{bak1989earthquakes,sornette1989self,olami1992self}. The virtual seismometer complex network shows power-law degree distribution with an exponent $\gamma=2.3\pm 0.2$. We also demonstrate that our model automatically predicts the seismic cycle or a period which depends on the time scales of the earthquake. The branching ratio inside and between avalanches reveal that the system is at (or more precisely close to) the critical point. We also reveal a considerable correlation between the dynamical Green's function and the nodes centrality.\\

The paper has been organized as follows: In the next section we describe some features of the Rigan earthquake. SEC.~\ref{network} devoted to a short introduction to the virtual seismometer analysis. In section~\ref{dynamics} we introduce the dynamical model and present the numerical details and results. We close the paper by a conclusion.

\section{Rigan earthquake}\label{Rigan}

\begin{figure*}
	\centerline{\includegraphics[scale=.5]{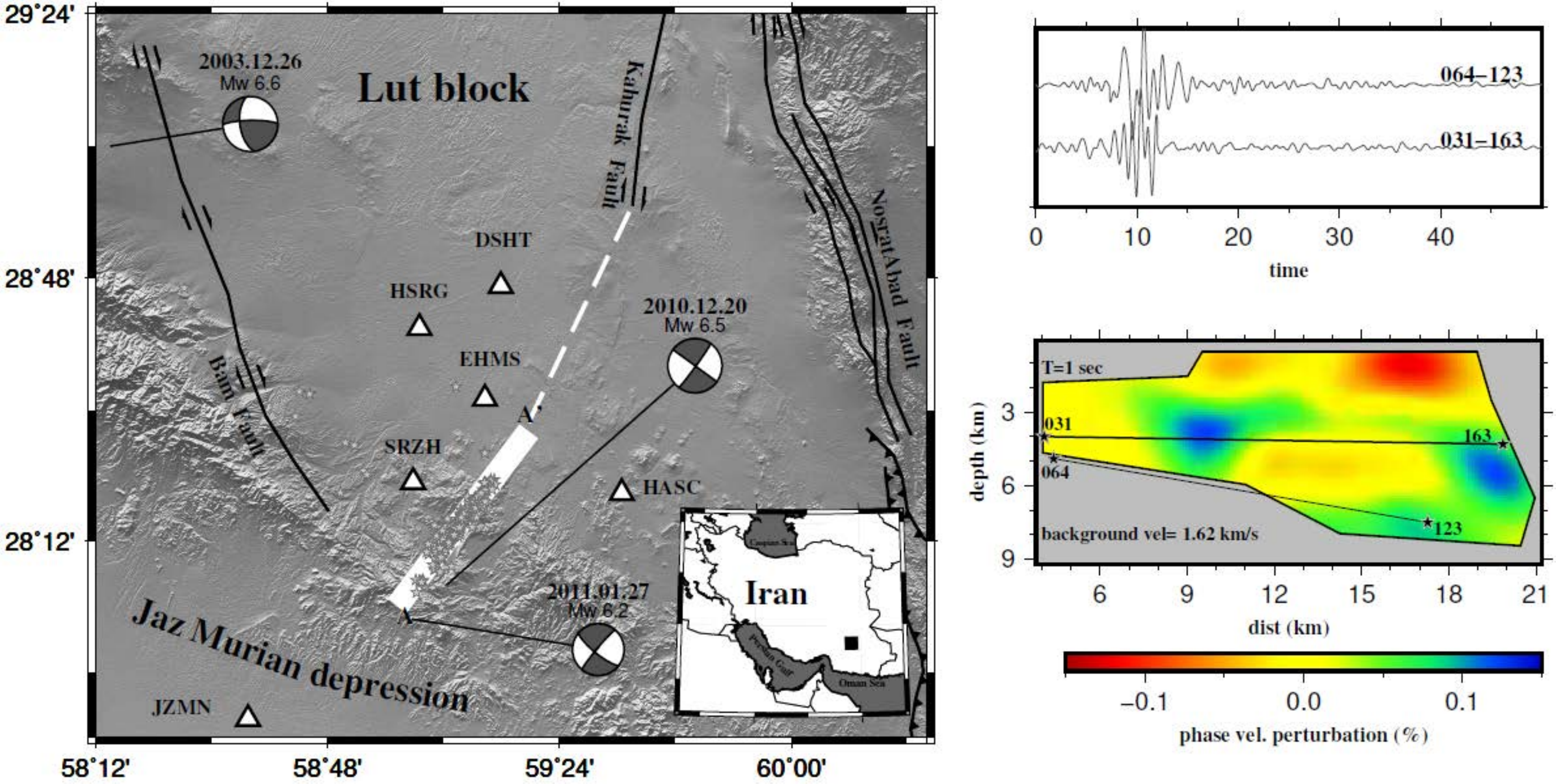}}
	\caption{(Color online): The general set up of the Rigan earthquake. The left figure is the map of the region of the earthquake, and the right figure shows the wave velocity field as the function of the position: from Ref.~\cite{shirzad2019}.}
	\label{fig:SeismicSheet}
\end{figure*}

The Rigan earthquake was occurred along the hidden part of the Kahurak Fault (see the dashed line in map of Fig.~\ref{fig:SeismicSheet}) with $Mw \ 6.5$ in the Kerman province of south Iran on 2010 December 20. The corresponding focal mechanism shows a right-lateral strike slip fault as depicted in Fig.~\ref{fig:SeismicSheet}. The Rigan area locates at the southern part of the Lut block and the northern edge of the JazMurian depression (Mirzaei, 1998). Bam Fault, Kahurak Fault, and NosratAbad Fault zone are the main fault systems which are surrounding this area (Fig.~\ref{fig:SeismicSheet}). These faults have been recently experienced a catastrophic earthquake (Bam earthquake occurred on 26 December 2003 with $Mw\ 6.6$~\cite{jackson2006seismotectonic}). In order to study the hidden part of Kahurak fault plane using aftershock events, a temporary network with six portable three-component stations (triangle in Fig.~\ref{fig:SeismicSheet}) was deployed by Iranian Seismological Center (IrSC) around the epicentral up to $40$ km radius. Recording continuous raw data was started on 2010 December 23, three days after the mainshock, until 2011 January 06. Rezapour and Mohsenpur (2013) by investigating this continuous data located $314$ aftershocks recorded by a minimum of $4$ stations with an azimuthal gap less than $180^{\deg}$, and with a root-mean-square (rms) of arrival time residual less than $0.2$ s. Some previous studies on this fault considered two individual and perpendicular faults around this mainshock (e.g.,~\cite{maleki2012relocation,walker20132010}). However, some of the other previous studies suggested a fault with rake angle in order of $80^{\deg}$ (\cite{rezapour20132010}),while most of aftershock events (more than 90$\%$) were occurred in a narrow band zone ($\sim 2$ km) along the defined escarpment of fault by~\cite{shirzad2013application}, so that the aftershocks' epicentral cover an area with $20\times 2$ km (the white box in map of Fig.~\ref{fig:SeismicSheet}). Since these aftershocks are related with steeper dip ($\sim 90^{\deg}$) of the fault with narrow band zone as shown using tomographic result by Shirzad~\cite{shirzad2013application}, the fault plane structure can be studied using appropriate tool. Shirzad et al.~\cite{shirzad2017near} combined ambient seismic noise and classical surface wave tomography to calculate radial anisotropy and crustal deformation. In spite of continuous data was recorded just for two weeks, they divided raw data to 10 minute window time and used a root-mean-square (RMS) stacking method to obtain inter-station EGFs with fairly high signal-to-noise ratio (SNR). Although this studies can give us an overview about rake angle (vary between $\sim 85-90^{\deg}$), subsurface layering, past and ongoing deformation in these layers around the fault, its resolution is not consummate in the depth greater than 5 km where the second mainshock occurred on 27 January 2011, approximately one month later. 
Curtis et al.~\cite{curtis2009} presented a virtual seismometer approach which can be applied to improve the resolution of surface wave tomography in regions with poor instrumental coverage. The Ref.~\cite{shirzad2019} then developed and used this method to obtain group and and phase semi-dispersion measurement models of this fault plane. In addition, the rotation matrix was applied for projecting semi-Rayleigh wave inter-event EGFs on fault plane in that study.

\section{The earthquake network}\label{network}

The complex networks dual to time series of seismic activities contain crucial information, containing the centrality of the activities, and the correlations between the events, all of which help a better understanding of its internal degrees of freedom, and the crucial mechanisms helping to predict the future activities. Constructing such complex dual system can be done in many ways. One of the most popular ones is the method described in~\cite{abe2006complex} in which a cubic grid of cells of equal size, covering the geographical zone of interest, is considered. A cell is considered as a node of the network if it contains a seismic event. Then a link is directed from node $A$ to $B$ if they occur successionally/sequentially in the time sequence of the earthquakes. In most cases the network that is constructed in this way is scale-free, with the power-law degree distribution $P(k)\sim k^{-\gamma}$, where $\gamma$ is the characteristic exponent. This exponent, although depends on the thresholds, usually has a value close to two for the real systems. For the $Mw=8.3$ Chile (Illapel-2015) earthquake, the $\gamma$ exponent was estimated to be $2.3\pm 0.2$ for the cell size $\Delta=10\ km$ and the magnitude threshold $M>3.0$~\cite{pasten2017time}. The visibility graph method is another choice which has been studied for many systems, like the seismicity of Italy between 2005 and 2010, yielding the $\gamma$ exponent from $1.9$ to $3.5$ depending on the threshold~\cite{telesca2012analysis}, and also the 2003-2012 earthquake sequence in the Kachchh Region of India~\cite{telesca2016visibility}.\\
These methods of constructing the complex networks based on the time sequence, although interesting, suffers a crucial problem: the connections in the dual network do not necessarily mean a physical connection between the events, i.e. two successive events are not necessarily related dynamically and they can simply occur in successive time order by accident, although the correlation between them cannot be excluded. By \textit{dynamically correlated} we mean that the activity of one of them induces the activity in the other. It is more demanding for us to find a method in which a connection is stablished for nodes that are dynamically correlated. For this, we introduce the virtual seismometer technique, using of which we are able to extract the spatial regions that are dynamically correlated.

\subsection{The virtual seismometer technique}\label{virtual}

Virtual seismometer approach, which is recently developed, could enable us to improve the resolution of surface wave tomography in the regions with poor instrumental coverage. In this method, Curtis et al.~\cite{curtis2009} suggested an alternative method to extract the inter-event EGF using the cross-correlating of event-pair waveforms recorded on a station. A simple description of this property represent as
\begin{equation}
\nabla_1\nabla_2G_h(x_2,x_1)=\sum_n\frac{u\left(x_{k_n}|x_2 \right)}{M_2}\frac{u\left(x_{k_n}|x_1 \right)}{M_1}e^{-j\omega_n\tau}
\label{Eq:Green}
\end{equation}
where the homogeneous Green's function and spatial derivative at $x_1$ and $x_2$ are indicated by $G_h$, $\nabla_1$ and $\nabla_2$ respectively. Also, $u$ and $M$ are displacement due to the Moment Tensor at $x_1$ and $x_2$. Moreover, the wavenumber and frequency define by $k$ and $w$. The relation~\ref{Eq:Green} can be summarized, a cross-correlation of an event-pair waveform is proportional to the inter-event EGF so that it is independent of source types (normal, thrust and/or strike-slip faults). Using this method, Shirzad (2019) obtained Rayleigh wave group and phase velocity models around fault plane at the periods of $1$, $2$ and $3$ sec. However, the Kahurak fault is the strike-slip type, a rotation matrix is applied for retrieving Rayleigh wave inter-event EGF signals. Compared to the previous interferometry studies (e.g.,~\cite{hong2006tomographic,tonegawa2009seismic}), some strict criteria were also applied by Shirzad~\cite{shirzad2019} in the data selection and/or preparation step which can be summarized as epicentral distance from fault escarpment less than $2.5$ km, the magnitude of aftershocks $M > 2.0$, horizontal and depth location uncertainties less than $2.5$ km, and the projection angle, $\Omega$, greater than $70$. Afterward, the cross-correlation of prepared waveforms has been done, and then phase weighing stacking procedure (see \cite{schimmel2011using}) has been performed to retrieve inter-event Rayleigh wave EGFs. An example of inter-event semi-Rayleigh wave EGFs depicts in the top panel  in right side of Fig.~\ref{fig:SeismicSheet}. For all retrieved inter-event EGF, group and  phase velocities semi-dispersion measurements has been calculated using Frequency-Time Analysis (\cite{herrmann1973some}) and zero crossing of Bessel function of the first kind (\cite{ekstrom2009determination}), respectively. In this step, the Gaussian filter, $\alpha$, coefficient is in order of 3 for multifilter technique, and average phase velocity semi-dispersion curves, which are separately calculated by the real and imaginary parts of the waveform in the frequency domain, has been applied. Moreover, many quality controls has been used on these retrieved inter-event EGFs for obtaining reliable tomographic maps based on Ref.~\cite{shirzad2015near} including (1) three times of wavelength less than epicentral distance ($3\lambda\leq \Delta$), (2) signal to noise of recorded waveforms greater than $4$, (SNR$_{\text{waveform}}$ $> 4.0$), and (3) group and/or phase velocity less than two times of standard deviation ($\sigma$) of observed velocities ($[V_{\text{average}}-2\sigma]\leq V\leq[V_{\text{average}}-2\sigma]$). Because of decreasing inter-event ray path, these criteria not only does not limit spatial resolution but also by identifying and then by rejecting all bad inter-event pathways ensures to obtain reliable tomographic results. Finally, the tomography procedure has been done by the iterative nonlinear inversion package of the Fast Marching Surface wave Tomography (FMST) which is developed by Rawlinson~\cite{rawlinson2005fmst}. This procedure is based on two main sub-steps, including the forward problem to obtain calculated travel time using the Fast Marching Method, and the inversion step to minimize between observed and calculated travel times. Inspections of Shirzad (2019) tomographic results (e.g., bottom panel in right side of Fig.~\ref{fig:SeismicSheet}) indicate a high velocity anomaly as a triangle shape in depth range of $3-6$ km. This result proposed this anomaly can be lead to trigger the second mainshock because \\
(1) this area surrounds with some of the aftershocks, which can be explained by asperity (see Aki 1984). \\
(2) The epicentral distance from the first mainshock is approximately $10$ km, which is matched with epicentral distances reported by Global CMT. \\
(3) The area of anomaly is approximately 15 km2 which can generate an earthquake with $MW$ $6.2$ for strike-slip faults as tabled by Well and Coppersmith (1994).\\
Moreover, the maximum distance from the fault escarpment and/or plane, where the tomographic results are reliable in this range, has been calculated using sensitivity kernel functions (see~\cite{shirzad2015near}) and it is up to $\sim 2$ km. Also, the stability of appeared anomalies has been investigated using different types of error and test resolutions.

\subsection{complex network of virtual seismometers}\label{network_results}

\begin{figure}
	\centerline{\includegraphics[scale=.65]{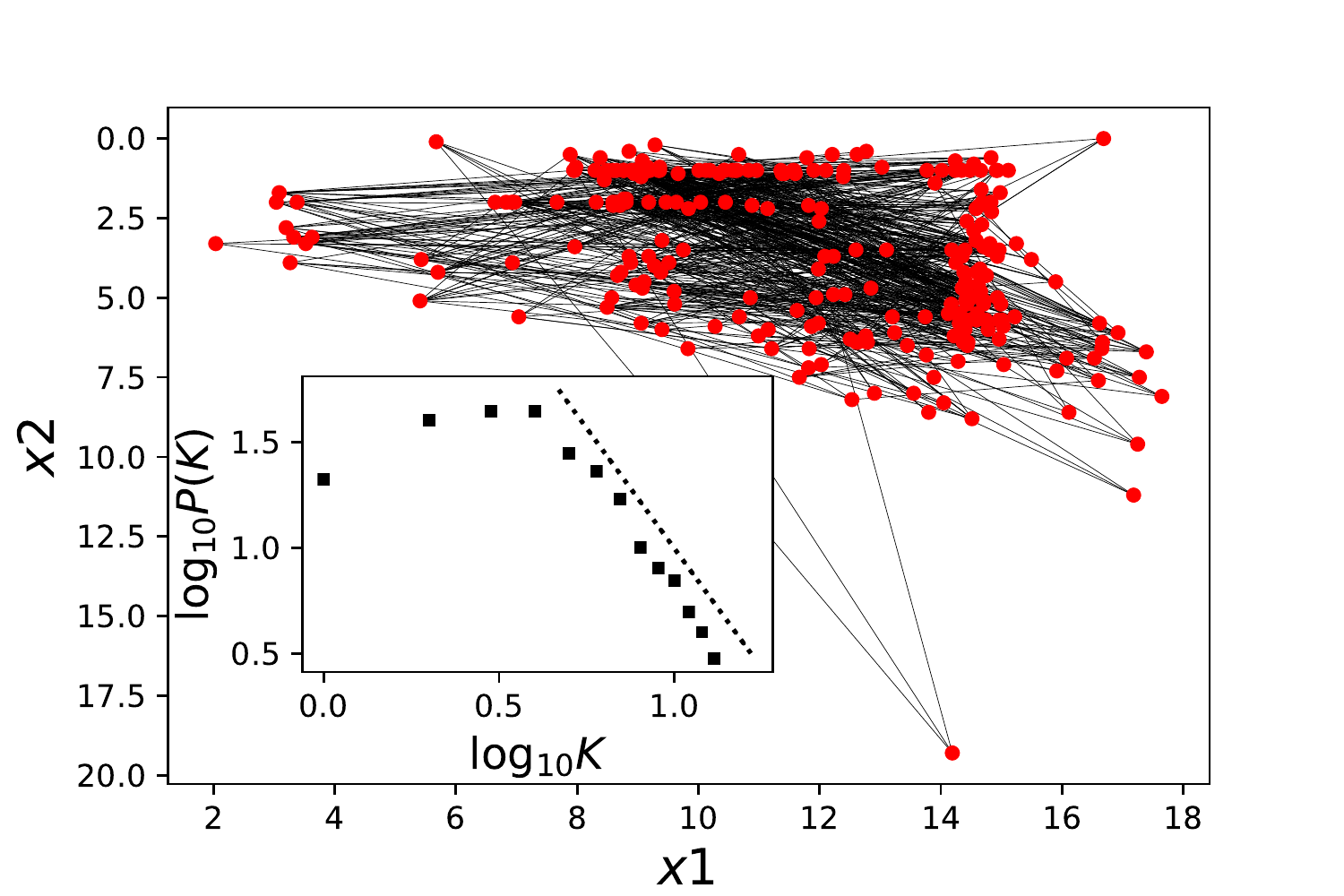}}
	\caption{(Color online): Undirected graph of earthquake nodes in two-dimensional Euclidean space (main panel) and the log–log plot of distribution of node degree ($K$) with exponent $\alpha = 2.3\pm0.2$ (inset).  }
	\label{fig:network}
\end{figure}

In this sub-section, we construct a complex network between hypocenter of aftershock events occurred on Kahurak fault plane, and study its properties. This network is constructed as connections between all event-pairs that the corresponding inter-event EGF signals have been extracted. In this study, each event, which has at least one connection with other events, is considered as a node. Let us show the connection and/or activity of $i$th node by $A_i(t)$ in which $t$ is time. Therefore, an edge (connection) between nodes i and j  is equivalent to the ray path coverage, which is used in the tomographic procedure. In the tomographic procedure, each ray path (connection) is defined by using observed/calculated group and/or phase velocity of each inter-event EGF. Therefore, the connections obtained by means of this method (calculating dynamical correlation) are by definition physical. The main earthquake was followed by 314 aftershock events. Using different data selection criteria, quality control in extracting inter-event EGF, the azimuthal direction and energy of extracted signals, the group and/or phase velocity constrain on the tomographic procedure, we decreased the total ray paths (connections) to $64$. After applying these conditions to construct network, we find an undirected connected graph shown in Fig.~\ref{fig:network} which is presented the ray path coverage on Kahurak fault plane. We found that the distribution of node degree $P(k)$ follows power-law relation. The corresponding exponent is obtained by means of the least square fit in the log-log plot to be $\gamma=2.3\pm 0.2$ (see the inset of Fig.~\ref{fig:network}). This value is compatible with the amount reported by~\cite{pasten2017time}, confirming that the virtual seismometer technique is reliable.\\

The next question is about the centrality of the graph to quantify the relevance of the regions (nodes in the network). This can be done by calculating the centrality of the graph, i.e. calculating the eigenvalues and eigenvectors of the adjacency matrix of the graph. We found that the largest eigenvalue of the graph (which is the most important quantity in the dynamical properties of the graph) is $\lambda=1.876$. If we consider a dynamics similar to that of Ref.~\cite{moosavi2017refractory} this means that the system is in the extended critical regime. The centrality field has been shown in the main panel of Fig.~\ref{fig:dynamical}, in which the inset shows the degree of nodes. The more colorful are the nodes, the more important they are. We also found very good correlation between the centrality, the degree, and the Green function (the latter being different from the Green function defined in Eq.~\ref{Eq:Green} will be defined in the next section). We see here that detecting the set of most important regions on the fault sheet which can be very helpful, become possible by means of virtual seismometer technique. Using the interferometry approach, an uncomplicated tool prepares to study a fault plane so that it can calculate the location and the magnitude (size) of a further earthquake on this fault plane (see Shirzad 2019). But, it cannot give us an overview of the time of occurrence of this earthquake because of the inherent problem.\\
Despite the fact that the above findings are very useful, we need some extra information concerning the size and energy and duration of the earthquakes taking place in this system, and also the (presumable) scaling relations between them. These do not come out from the virtual seismometer analysis, since the data is quite small. In fact, it is the aim of the present paper to predict the seismic behavior of a system with small set of data. To this end, we switch to simulations, and consider a dynamical model on top of the system. Although the analyzed graph has been obtained by means of the dynamics of seismic system, it yields the correlations between regions, i.e. it tells us a part of system effectively affects which part/is affected from which part of the system. Therefore the graph is a very good candidate to be host of a relevant dynamical system capturing the physics of earthquake. This dynamical model, whatever it is, should contain the following requirements:\\
1- The local relaxation of stress (as the relevant field for earthquake),\\
2- It should define some local stress thresholds under which the plate is (locally) static.\\
A very good candidate for this model is sandpile model introduced by Bak, Tang, and Weisenfeld (BTW)~\cite{Bak1987Self}. This is done in the next section.


\section{Dynaimcal aspects}\label{dynamics}
The physics of the fault sheets dynamics is the center of attention for physicists and seismologists, for which many statistical models have been introduced. The spring-block model is one of them in which the spring strain spreads throughout the system by means of avalanche-like dynamics, and in each local activity the stress is distributed isotropically between the closest neighbors~\cite{burridge1967model,brown1991simplified}. These systems interestingly organize themselves in critical state~\cite{bak1989earthquakes}, which inspired many studies based on cellular automata model~\cite{olami1992self}. In this section we consider the BTW model on top of the graph that we obtained in the previous section.

\subsection{The model and simulation method}\label{BTW}

\begin{figure*}
	\centering
	\begin{subfigure}{0.49\textwidth}\includegraphics[width=\textwidth]{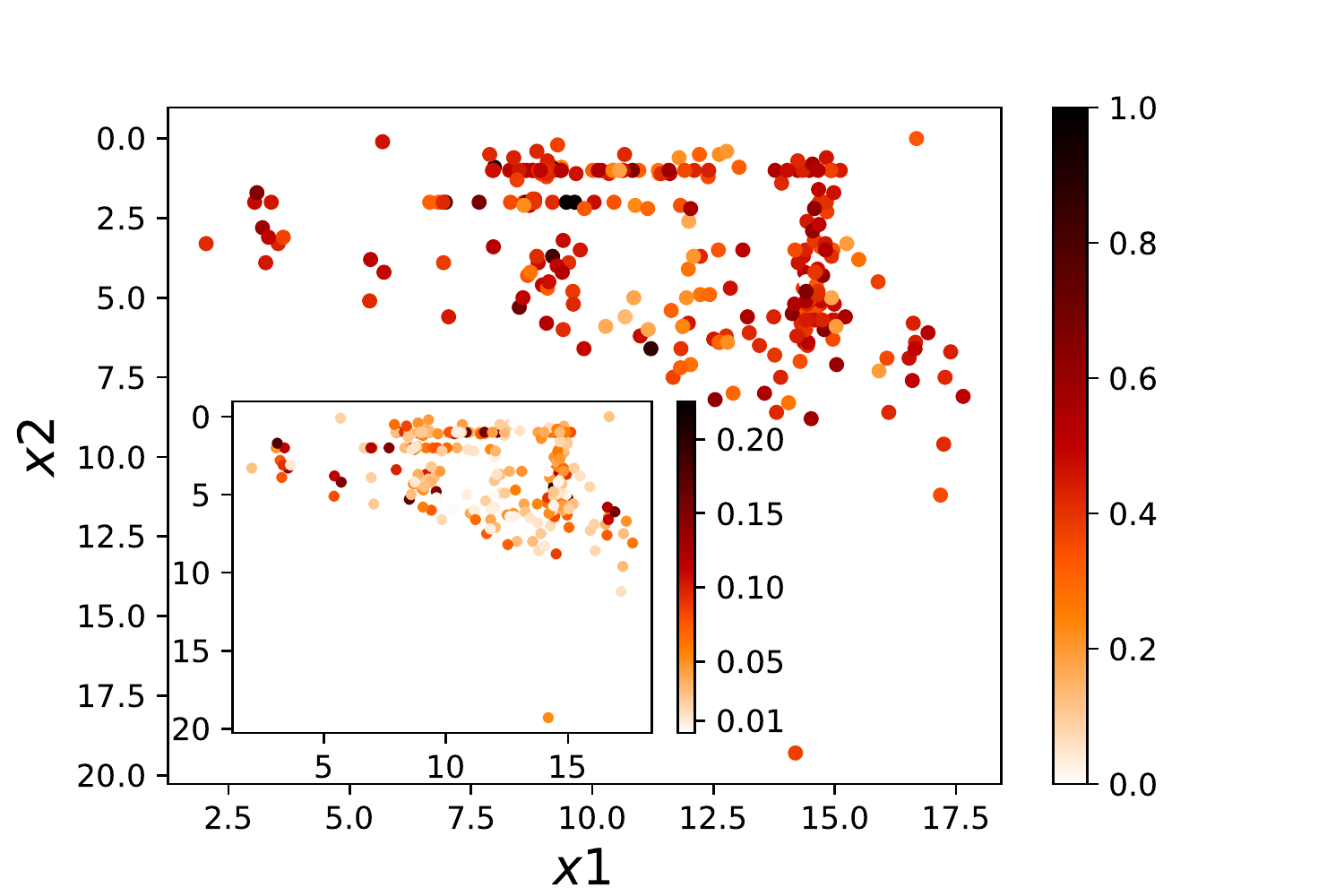}
		\caption{}
		\label{fig:Green_Function}
	\end{subfigure}
	\begin{subfigure}{0.49\textwidth}\includegraphics[width=\textwidth]{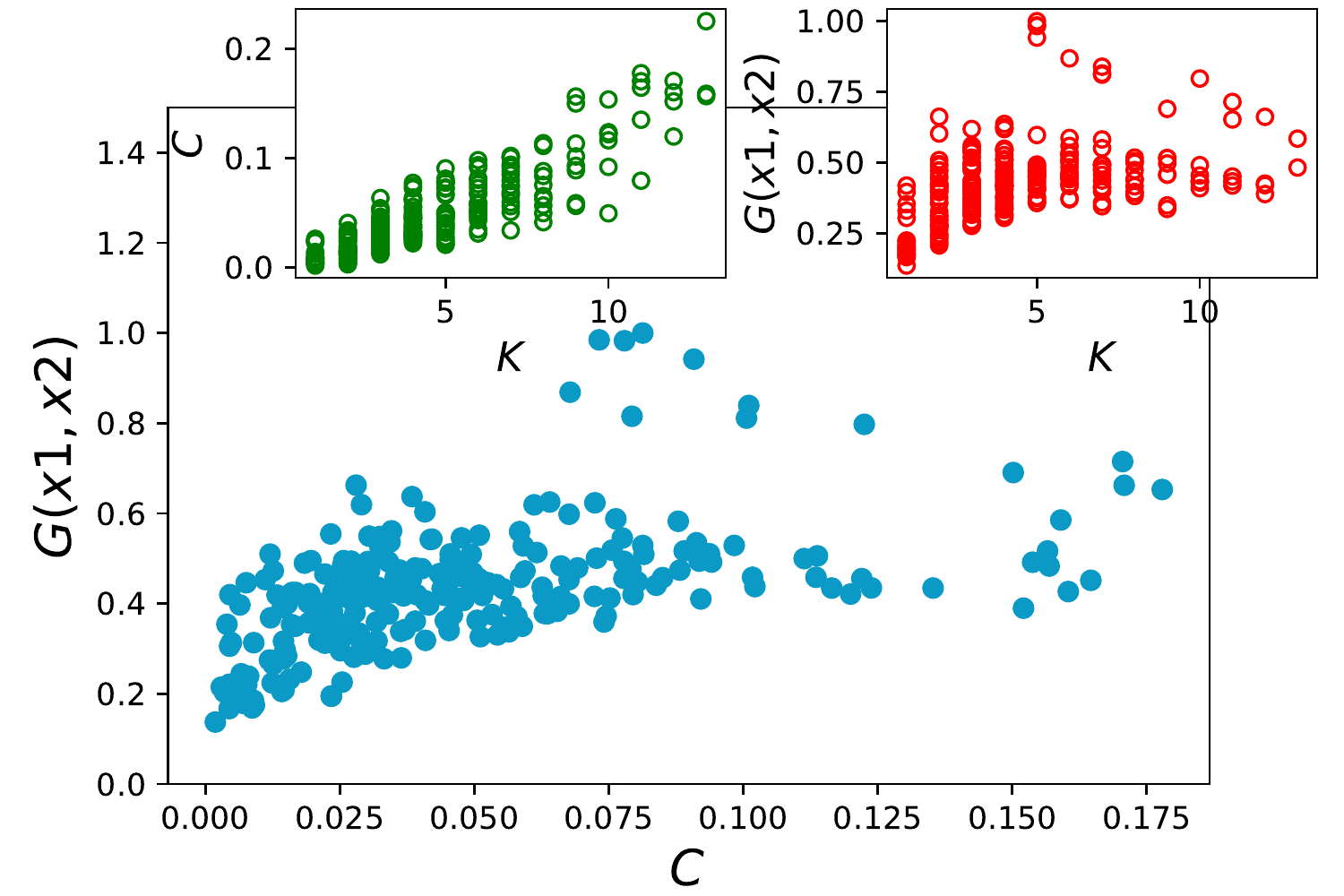}
		\caption{}
		\label{fig:Cent_Deg_Green}
	\end{subfigure}
	\caption{(Color online): (a) Green function of nodes when stimulated only the node with largest eigenvector centrality with coordinates $(14.4,4.5)$. inset of (a): the eigenvector centrality of each node when largest eigenvalue of network is equal $1.876$. (b) relation between centrality($C$) and Green function (main panel), between node degree ($k$) and centrality (left inset) and between node degree and green function (right inset).}
	\label{fig:dynamical}
\end{figure*}

Let us suppose that the stress (energy) units spread throughout the nodes of the graph (obtained in the previous section) according to the BTW-type dynamics. Put in other words, we consider the BTW dynamical model on top of the graph. In the BTW dynamics, we consider on each node $i$ a stress (or a local energy) $\epsilon_i$ (the number of sand grains) taking initially their values randomly (independently and uncorrelated) with the same probability one integer from $\left\lbrace 1,...,Z_i\right\rbrace $, in which $Z_i$ is the number of the nodes connected to the node $i$, i.e. the degree of $i$. Then we stimulate a random site $i$ by increasing its local stress, so that $\epsilon_i\rightarrow \epsilon_i+1$ (note that $1$ has arbitrarily been chosen as the stress unit). As a result this site may become \textit{unstable} ($\epsilon_i>\epsilon^{\text{th}}\equiv Z_i$), which cause a local relaxation process to start, during which $\epsilon_j\rightarrow \epsilon_j-\Delta_{i,j}$, where
\begin{equation}
\begin{split}
\Delta_{i,j}=\left\lbrace \begin{matrix}
-1 & \text{if}\ i \ \text{and}\ j \ \text{are neighbors}\\
Z_i & \text{if}\ i=j\\
0 & \text{otherwise}
\end{matrix}\right. 
\end{split}
\end{equation}
After a node relaxes, it may cause the neighbors to become unstable and relax, and so on, continuing until no node is unstable anymore. Then another random site is chosen for stimulation and so on. The stress can be dissipated from a sink node defined as a node with no outgoing link. The average height grows with time, until it reaches a stationary state after which the total stress (defined as the summation of all local stresses) is statistically constant (surely with some fluctuations). The dynamics can be implemented with either sequential or parallel updating. Let us parametrize a single avalanche by the internal time $t'$ in such a way that when $N$ search is performed for the unstable sites (to be toppled), then $t'\rightarrow t'+1$. Then the number of nodes that are relaxed at $t'$ is denoted by $s(t')$, and $T$ defined as the maximum of $t'$ is the duration of the avalanche. An avalanche is defined as the process that is started by a single stimulation, and is ended when no node is unstable.\\
To study the interplay between the dynamical model, and the host graph, we have calculated the Green function defined as follows: suppose that the node $i$ is stimulated causing an avalanche. Then the Green function $G(i,j)$ is the number of times that node $j$ relaxes. For calculating this function, we have stimulated only the node with largest outgoing links, and calculated $G(x_1,x_2)$ as the number of times that the node located at $(x1,x2)$ relaxes. This is shown in the main panel of Fig.~\ref{fig:Green_Function}, in which the inset shows the centrality. The relation of between the Green function, the centrality and the node degrees is shown in Fig.~\ref{fig:Cent_Deg_Green}, in which each point show the obtained values for the e.g. the Green function and the centrality of a node. For the main panel, although the points are scattered, but they are gathered in a region, showing that the Green function and the centrality are correlated.\\

\begin{figure}
	\centerline{\includegraphics[scale=.5]{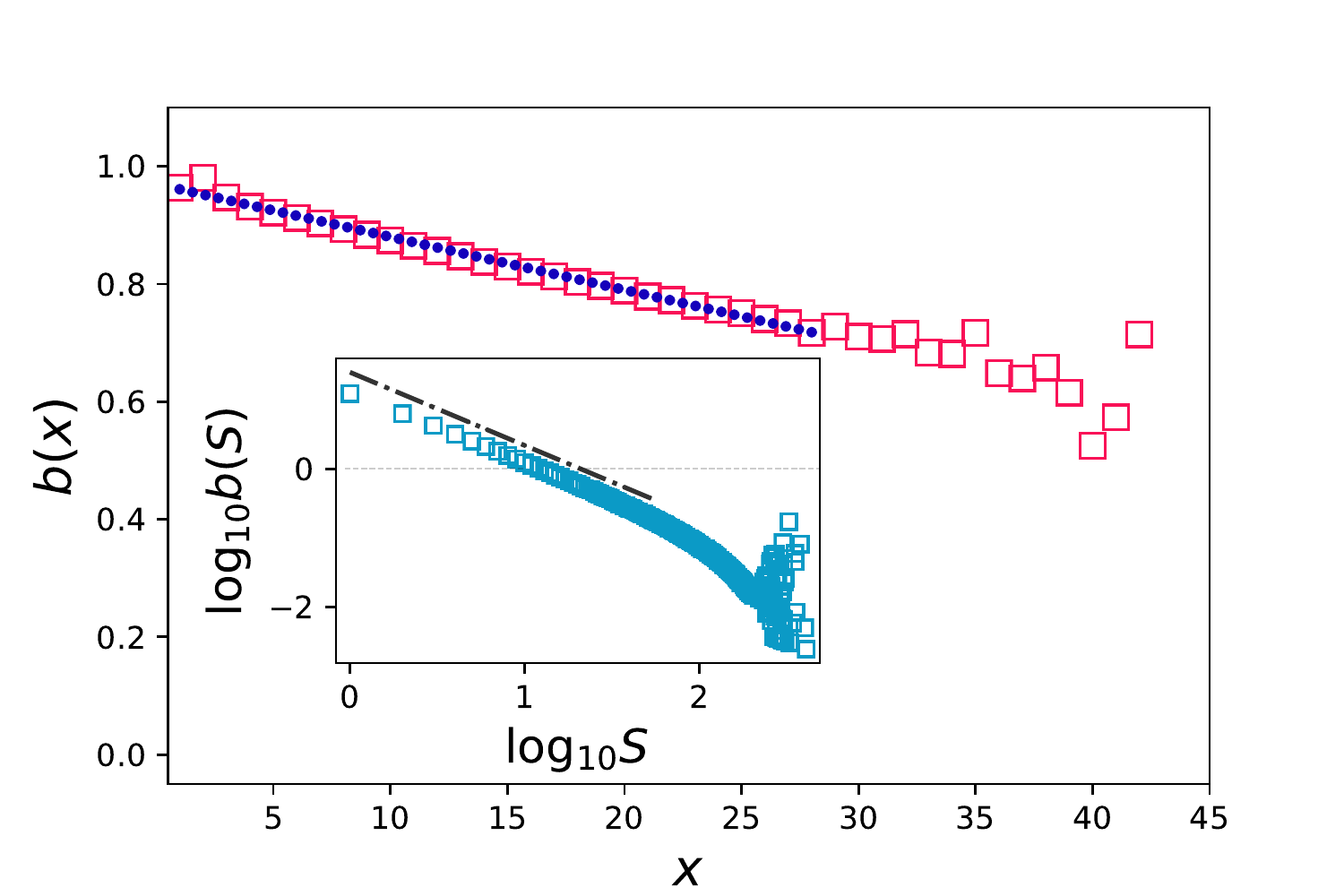}}
	\caption{(Color online): Activity dependent branching ratio $b(x)$ for instantaneous avalanche sizes; blue line represent it's trend line with equation $b(x)=-0.009x+m;\ m=0.970\pm0.006$ (main panel) and log-log plot of branching ratio for total avalanche size with exponent $\alpha =1.07\pm0.01$ (inset).}
	\label{fig:b}
\end{figure}

The quantities that are of especial importance are the avalanche duration $T$, the avalanche size $S\equiv \sum_{t'=1}^{T}s(t')$, and the avalanche mass $M$ which is defined as the number of nodes that at least are triggered once during the avalanche. Also the criticality of the system can be tested using the branching ratio function $b(X)$ defined by
\begin{equation}
b(X)=E\left[\left. \frac{s(t'+1)}{X}\right| s(t')=X \right] 
\end{equation}
where $E[A|B]$ is the conditional expectation value (ensemble average) of $A$ conditioned that $B$ is already satisfied, and $X(t')$ is the number of unstable nodes at the internal time $t'$. It is shown that for the criticality $\lim_{X\rightarrow 0}b(X)=1$~\cite{alstrom1988mean}. This function is shown in the main panel of Fig.~\ref{fig:b}. We see that $b(X)$ increase linearly as $X$ decreases towards the final value $0.98\pm 0.02$ at $X=0$. This confirms that the system is in the critical state. Also in the inset we show $b(S)$ which has a same definition, but here for $S(t)$ where $t$ is the external time defined as the avalanche number. We see that $b(S)$ behave in power-law form, with the solution the solution of $b(S^*)=1$ is $S^*=10.0\pm 0.5$, which is the fixed point of the dynamics. Therefore, on the mean field level this system is active meaning that, in the existence of the external stress support, it organizes itself in a critical state with a mean avalanche size $S^*$.

\begin{figure*}
	\centering
	\begin{subfigure}{0.49\textwidth}\includegraphics[width=\textwidth]{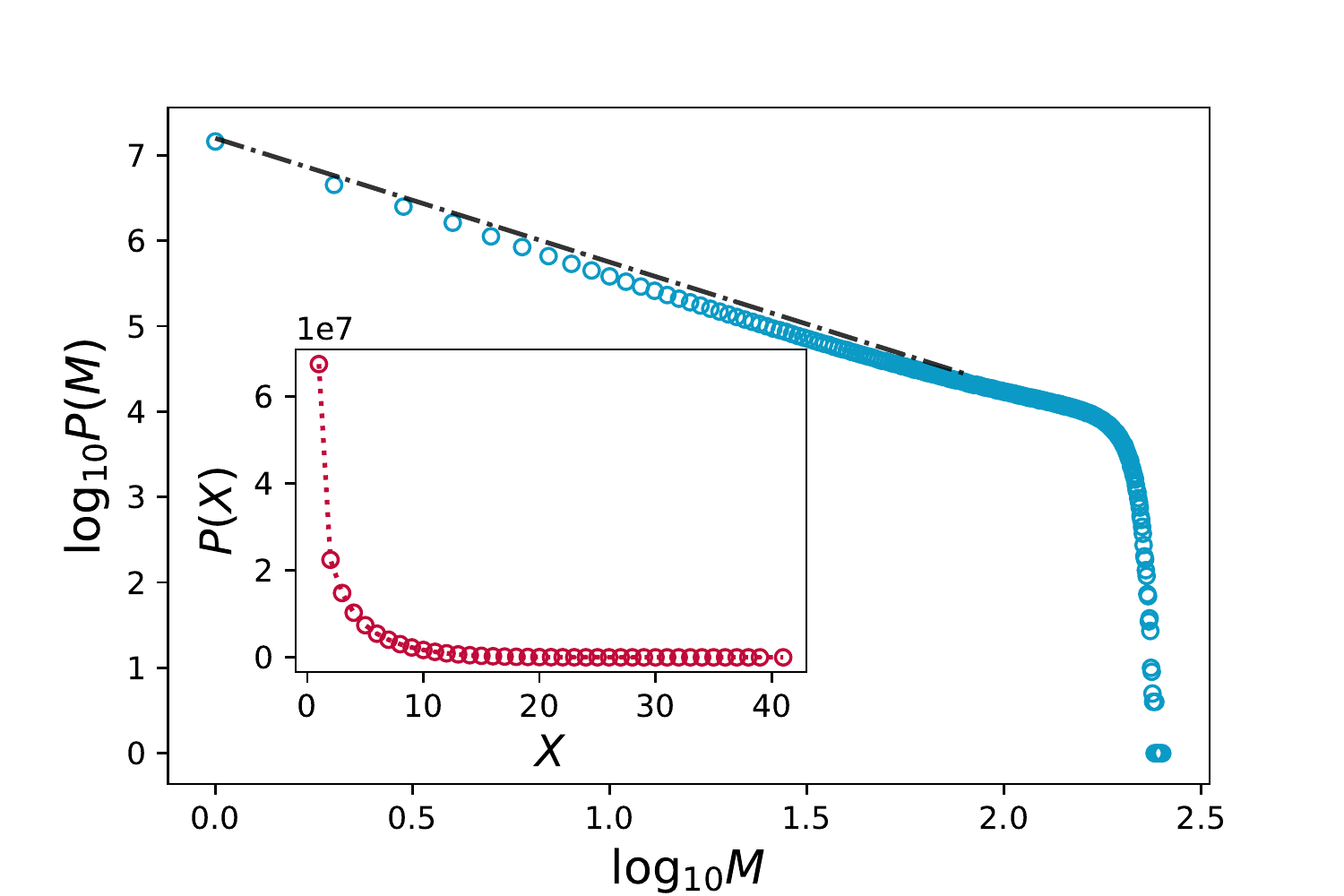}
		\caption{}
		\label{fig:p(t)}
	\end{subfigure}
	\begin{subfigure}{0.49\textwidth}\includegraphics[width=\textwidth]{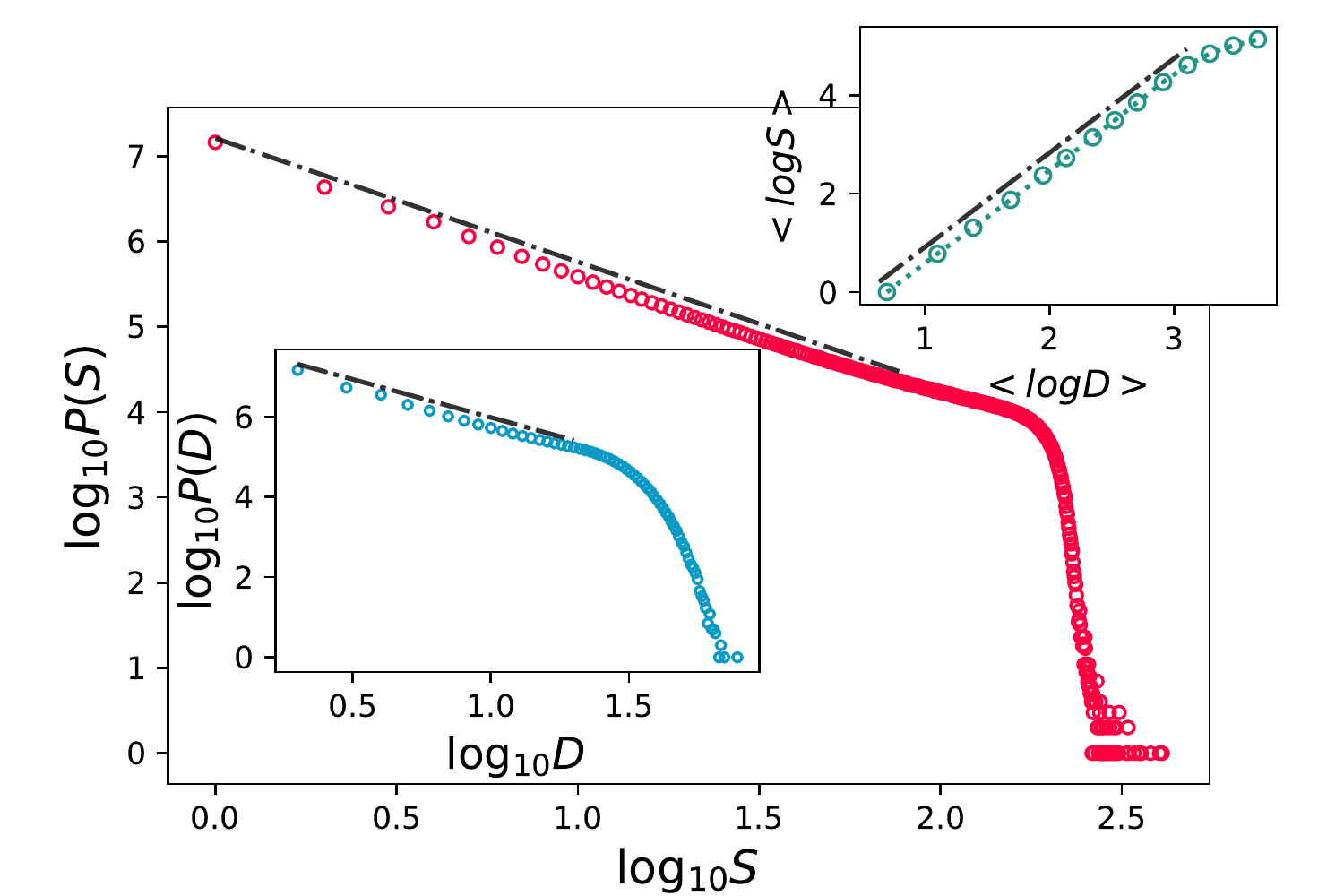}
		\caption{}
		\label{fig:p(x)}
	\end{subfigure}
	\caption{(Color online): (a) The $log-–log$ plot of the distribution functions of avalanche mass $M$ with exponent $\alpha =1.44\pm0.02$ (main panel) and the distribution functions of the instantaneous avalanche sizes $x$ (inset). (b) The $log-–log$ plot of the distribution functions of avalanche size $S$ with exponent $\alpha =1.45\pm0.02$. Lower insets: the same graph for the distribution functions of avalanche duration $D$ with exponent $\alpha =1.90\pm0.07$. Upper insets: The $log-–log$ plot of $S-D$ diagram with exponent $\alpha =1.92\pm0.02$.}
\label{fig:exponents}
\end{figure*}

For the (here self-organized) critical systems some power-law behaviors appear, defining some critical exponents. For example the distribution function of the variable $x$ shows scaling relation (for the infinite system) $P(x)\sim x^{-\tau_x}$ where $x=S,T,M$, and $\tau_x$ is their exponent. For the finite systems, the power-law dependence is destroyed at some point in which the finite size effects become important. Additionally some scaling relations are commonly seen between the variables, i.e. $y\sim x^{\gamma_{yx}}$ where again $x,y=S,T,M$, and $\gamma_{yx}$ is the corresponding exponent. By a simple calculation, one can show that a following hyper-scaling relation should hold:
\begin{equation}
\gamma_{ST}^{\text{hyperscaling}}=\frac{\tau_T-1}{\tau_S-1}
\end{equation}
We have shown the distribution functions for $T$ (Fig.~\ref{fig:p(t)}), $S$ (main panle of Fig.~\ref{fig:p(x)}), and $M$ (inset of Fig.~\ref{fig:p(x)}). Using the least square fit of the linear part of the log-log plot, we obtain that the exponents are $\tau_T=1.90\pm 0.07$, $\tau_S=1.45\pm 0.02$, and $\tau_M=1.44\pm 0.02$. We see that the exponents are very close to the mean filed exponents ($\tau_S^{\text{mean field}}=\tau_M^{\text{mean field}}\approx\frac{3}{2}$~\cite{chessa1998mean}). Due to largely scattered values for the reported exponents in various studies~\cite{olami1992self} on the natural systems, we cannot judge about the obtained exponents.  Using these value one obtains $\gamma_{ST}^{\text{hyperscaling}}=2.00\pm 0.18$, which should be compared with $\gamma=1.92\pm 0.02$ (inset of Fig.~\ref{fig:p(t)}), showing that the hyperscaling relation holds.\\
 
\begin{figure}
	\centerline{\includegraphics[scale=.55]{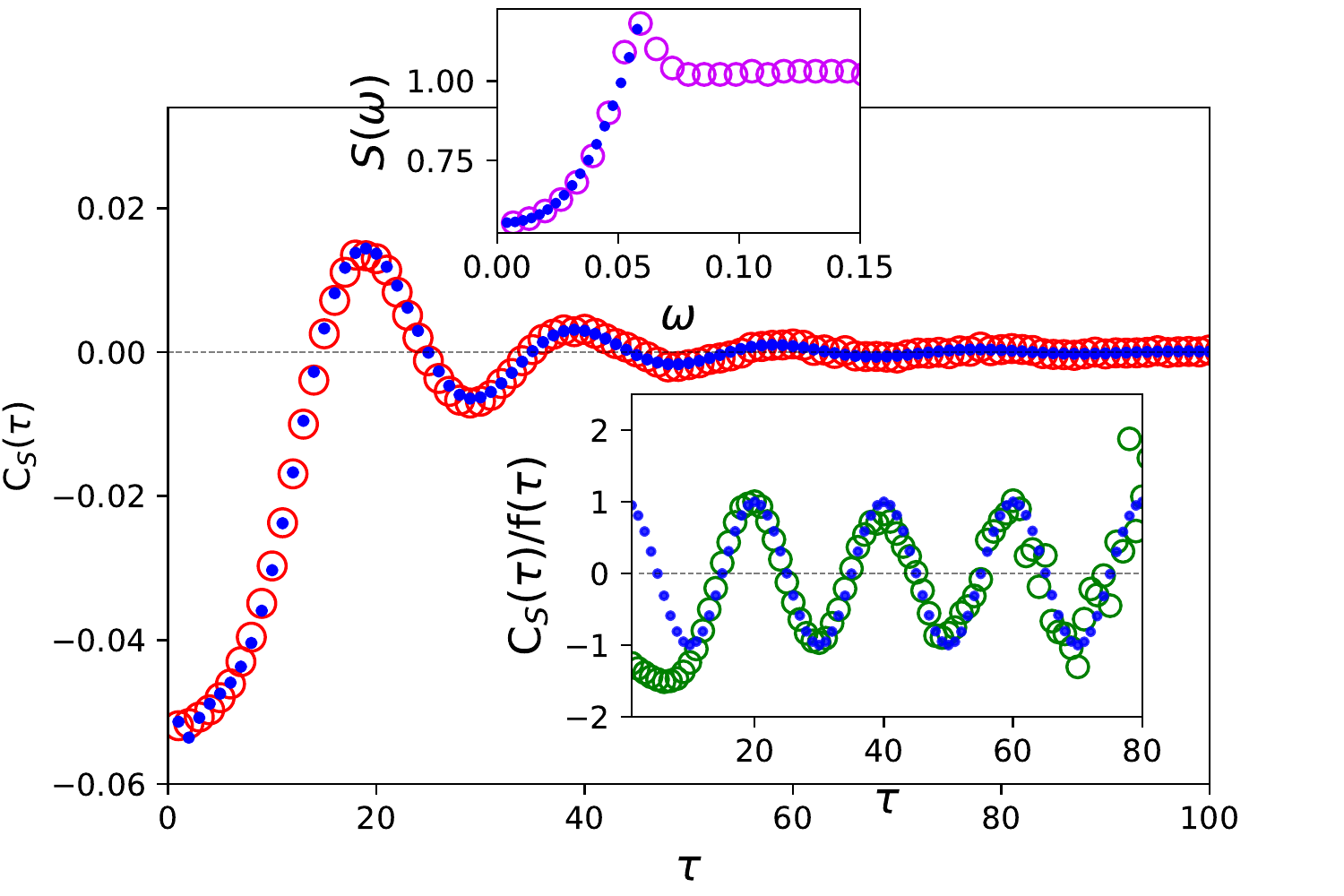}}
	\caption{(Color online):  Auto-correlation function $C_\tau$ for total size of avalanches S; blue line represent it's trend line with equation $C_s(\tau)=f_0e^{-\alpha_1\tau}\left[\cos \left( 2\pi\frac{\tau}{\tau_1}\right)-\frac{f_1}{\tau^{\alpha_2}}\cos \left( 2\pi\frac{\tau-\tau_0}{\tau_2}\right)\right]$ (main panel). Lower insets: the same graph witch Auto-correlation function divided by $f(\tau)=f_0e^{-\alpha_1\tau}$; blue line represent it's trend line with equation $C_s(\tau)/f(\tau)=\cos(\frac{2\pi}{\tau_1}\tau)$. Upper inset: Power spectrum of total size of avalanches S; blue line represent it's trend line with equation $S_{0}(\omega)+A\omega^{\alpha};\ \alpha=2.6\pm0.1.$ }
	\label{fig:Power_Spectrum}
\end{figure}

An interesting quantity for characterizing the avalanches is the auto-correlation function $C_{\tau}$ between distinct avalanches, and is very helpful in realizing their structure. If we consider the time series of $S(t)$ (here $t$ is the avalanche number), then it is defined as:
\begin{equation}
C_{\tau}\equiv \frac{\left\langle S(t+\tau)S(t)\right\rangle - \left\langle S(t)\right\rangle^2 }{\left\langle S(t)^2\right\rangle - \left\langle S(t)\right\rangle^2}
\end{equation}
For the BTW model on regular 2D lattice, this function is zero for $\tau\ne 0$ for the waves, whereas it is long-range for avalanches, signaling that they are not mono-fractal. For our case, it is shown in Fig.~\ref{fig:Auto_correlatin_Power} (note that $C_{0}=1$, which is not shown in the figure), from which we see an interesting oscillatory behavior. In the branching ratio analysis, we saw that the fixed point $S^*$ is uniquely determined, which shows that the average stress in the stationary state is single, i.e. the system is not in the oscillatory phase and the oscillatory behavior of the auto-correlation function has other sources. By fitting this function, we notice that it has two harmonic components, one of which decays in a power-law fashion with time:
\begin{equation}
C_s(\tau)=f_0e^{-\alpha_1\tau}\left[\cos \left( 2\pi\frac{\tau}{\tau_1}\right)-\frac{f_1}{\tau^{\alpha_2}}\cos \left( 2\pi\frac{\tau-\tau_0}{\tau_2}\right)\right] 
\end{equation}
where $f_0=0.030\pm0.003$, $\alpha_1=0.057\pm0.004$, $\tau_1=19.6\pm0.1$, $\tau_2=27.04\pm0.53$, $f_1=17.38\pm1.2$, $\alpha_2=1.26\pm0.04$, $\tau_0=7.08\pm0.13$. In the inset $\frac{C_s(\tau)}{f_0e^{-\alpha_1 \tau}}$ is shown to show more evidently the oscillatory behavior. This oscillatory behavior should cause a peak in the power spectrum, as shown in Fig.~\ref{fig:Power_Spectrum}. The peak of this function is at $\bar{\omega}$ which is consistent with $\tau_1$, i.e. $\bar{\omega}=\frac{2\pi}{\tau_1}$ as expected. \\

In addition to the oscillatory behavior, this finding shows that there is a refractory period $\frac{\tau_1}{2}$, so that when a large avalanche takes place, up to this time the probability of having a large event is small. The interesting fact is that after time $\tau_1$, the probability of having a large event is maximal. In the Rigan earthquake, after $20$ days another earthquake took place, which we think that it should be explained by this observation on $C_s(\tau)$.
\begin{table}
	\begin{tabular}{c | c | c }
		\hline exponent & definition & value \\
		\hline $\tau_S$ & $P(S)\sim S^{-\tau_S}$ & $1.45\pm 0.02$ \\
		\hline $\tau_T$ & $P(T)\sim T^{-\tau_T}$ & $1.90\pm 0.07$ \\
		\hline $\tau_M$ & $P(M)\sim M^{-\tau_M}$ & $1.44\pm 0.02$ \\
		\hline $\gamma_{ST}$ & $S\sim T^{\gamma_{ST}}$ & $1.92\pm 0.02$ \\
		\hline $\tau_1$ & fit in Fig.\ref{fig:Power_Spectrum} & {$19.6\pm0.1$} \\
		\hline
	\end{tabular}
	\caption{The exponents for the .}
	\label{tab:exponents}
\end{table}

\section*{Discussion and Conclusion}
\label{sec:conc}

This paper is devoted to the statistical analysis of the Rigan earthquake 2010 December 20. For this, we used the recent technique designed by Curtis et al.~\cite{curtis2009}, and developed further by Shirzad~\cite{shirzad2019}, according to which one extracts the inter-event EGF using the cross-correlating of event-pair waveforms recorded on a real station. Using this method, and by defining some thresholds/criteria, we obtained a complex network, whose node's degree distribution is found to be power-law with an exponent $\gamma=2.3\pm 0.2$, consistent with the range of exponents that were found in other earthquake (e.g.~\cite{pasten2017time}). After extracting this network, we implemented a dynamical avalanche model similar to the BTW-sandpile model on top of this network which has already proved to be acceptable for estimating the behavior of the avalanches~\cite{bak1989earthquakes,sornette1989self,olami1992self}. The numerical calculation of branching ratio demonstrated that the system is in the critical state with power-law behavior for the distribution function of the size and the mass and the duration of the avalanches, and with some scaling relations between these quantities. The critical exponents (and their definitions) are presented in TABLE~\ref{tab:exponents}. Also a strong correlation between the dynamical Green function and the nodes centralities has been observed, demonstrating a correlation between the dynamical model and the support complex network.\\

By calculating the dynamical auto-correlations of the avalanches, we show that this model yields naturally the seismic cycle found already in earthquakes, which translates to intermittency. These functions are composed of two decaying periodic terms with nearly the same period, one of which decays further in a power-law fashion that is killed for long enough times (Fig.~\ref{fig:Power_Spectrum}). This function realizes the \textit{seismic cycle} in real earthquakes during which a large event is followed by a shadow period of quiescence and then a new approach back toward the critical state, in which the events become larger~\cite{jaume1999evolving,huang1998precursors,sammis1999seismic}.

\bibliography{refs}

\end{document}